\documentclass[11pt,a4paper]{article}
\usepackage{geometry}
\usepackage{graphics}
\usepackage{setspace}
\usepackage[scaled=0.9]{helvet}
\usepackage{graphicx}
\usepackage{subfigure}
\usepackage{amssymb}
\usepackage{rotating}
\usepackage{arydshln}
\usepackage{multicol}
\usepackage{abstract}
\usepackage{ucs}
\usepackage[utf8x,utf8]{inputenc}
\usepackage[T1]{fontenc}
\usepackage[swedish, english]{babel}
\usepackage{fancyhdr}

\begin{document}
\bibliographystyle{acm} 
\pagestyle{fancy}
\cfoot{\thepage}
\renewcommand{\abstractname}{}

\title{\fontfamily{phv}\selectfont{\huge{\bfseries{A population-based approach to background discrimination in particle physics}}}}
\author{
{\fontfamily{ptm}\selectfont{\large{Federico Colecchia$^{1,2}$}}}\thanks{Email: federico.colecchia@brunel.ac.uk}\\\\
{\fontfamily{ptm}\selectfont{\large{{\it $^1$ Department of Physics and Astronomy, University College London}}}}\\
{\fontfamily{ptm}\selectfont{\large{{\it Gower Street, London WC1E~6BT, UNITED KINGDOM}}}}\\\\
{\fontfamily{ptm}\selectfont{\large{{\it $^2$ Brunel University}}}}\\
{\fontfamily{ptm}\selectfont{\large{{\it Kingston Lane, Uxbridge, Middlesex UB8 3PH, UNITED KINGDOM}}}}
}
\date{}
\maketitle
\begin{onecolabstract}
Background properties in experimental particle physics are typically estimated using control samples corresponding to large numbers of events. This can provide precise knowledge of average background distributions, but typically does not consider the effect of fluctuations in a data set of interest. A novel approach based on mixture model decomposition is presented as a way to estimate the effect of fluctuations on the shapes of probability distributions in a given data set, with a view to improving on the knowledge of background distributions obtained from control samples. Events are treated as heterogeneous populations comprising particles originating from different processes, and individual particles are mapped to a process of interest on a probabilistic basis. The proposed approach makes it possible to extract from the data information about the effect of fluctuations that would otherwise be lost using traditional methods based on high-statistics control samples. A feasibility study on Monte Carlo is presented, together with a comparison with existing techniques. Finally, the prospects for the development of tools for intensive offline analysis of individual events at the Large Hadron Collider are discussed.
\end{onecolabstract}

\begin{multicols}{2}

\section{Introduction}
\label{intro}
Background discrimination in particle physics is usually performed by identifying events that are more likely to contain a physics process of interest, the primary goal being rejection of contributions from uninteresting processes that mimic the signal and thus make its extraction and measurement more complicated. Traditional approaches achieve this goal by focusing on entire events, comparing kinematic and topological properties with reference distributions usually obtained from control samples.

This article presents a novel approach that builds on a population-based view of particle physics events. Events are treated as mixtures of subpopulations comprising particles originating from different physics processes such as a hard scattering of interest as opposed to background 
associated with low-energy strong interactions. The main goal is to decompose an input data set by assigning individual particles a probability for them to originate from a given process based on particle-level information.

This is achieved by adapting and applying mixture decomposition techniques \cite{marin} that are well established in statistics and that have been used in other disciplines to solve formally-similar problems. In this formulation, events are treated as heterogeneous statistical populations comprising particles whose kinematics reflects the process they originated from.

This contribution describes an initial investigation of the possibility to use mixture model decomposition techniques for background discrimination at the Large Hadron Collider (LHC). The study is based on a sampling algorithm inspired by the Gibbs sampler \cite{geman} and by Expectation Maximization (EM) \cite{EM} that decomposes an input data set into collections of particles originating from a hard scattering of interest as opposed to background 
associated with low-energy strong interactions, mapping individual particles to signal or background on a probabilistic basis. A number of well-established methods and results set a context for this investigation in addition to the Gibbs sampler and to EM, namely (i) other simulation-based methods such as the one documented in \cite{TW}, (ii) a more general use of Markov Chain Monte Carlo (MCMC) techniques, recently applied to the study of the Cosmic Microwave Background radiation \cite{CMB}, (iii) a recent renewed interest in Bayesian numerical methods for data analysis in particle physics \cite{dagostini} \cite{dagostini2} \cite{ciuchini}, in addition to (iv) the use of MCMC with reference to specific optimization problems in the field \cite{hepdata}.

In this study, the proposed sampling algorithm was used to classify individual particles into signal and background. The results obtained on a collection of $\sim 600$ simulated particles from a hard scattering and from background associated with low-energy strong interactions are presented and discussed, together with cross-checks on toy Monte Carlo as described in the appendix.


In general, different events in particle physics can look very different from one another even when the underlying physics processes are the same, and the effect of fluctuations can be non-negligible in low-statistics data sets. If we were to classify particles inside events into signal or background using traditional supervised methods, fluctuations would not be taken into account. In fact, training typically relies on high-statistics control samples where the effects of fluctuations are normally washed out. On the other hand, the algorithm presented in this article can estimate the shapes of signal and background particle-level probability density functions (PDFs) from the data: this makes it possible to use information extracted from the data to improve on the description of the signal and background PDFs obtained from control samples. 

From a broader perspective, this contribution illustrates a new population-based approach that aims to improve on the description of background PDFs obtained from a high-statistics control sample by estimating the effect of fluctuations in a data set of interest. This is done by assigning individual particles a probability for them to originate from signal or background, i.e. by decomposing an input collection of particles into a signal and a background-associated subpopulation.


\section{The algorithm}
\label{GS}

This approach to background discrimination is presented with reference to the general problem of decomposing a collection of particles from high-energy particle collisions into subpopulations associated with different underlying physics processes and described in terms of different PDFs.

The input data set consists of a mixture of particles, some of which originated from a hard scattering of interest, others from background associated with low-energy strong interactions. Provided that the corresponding subpopulations can be characterized sufficiently well in terms of their kinematic or topological properties, it is possible to ask, for each particle, what the probability is for it to originate from signal as opposed to background. In particular, the proposed algorithm estimates such probabilities by iteratively sampling from subpopulation PDFs.


As opposed to classical mixture models, which typically rely on a parametric formulation requiring the shapes of the subpopulation PDFs to be known a priori, our formulation is based on a more general mixture of the form

\begin{equation}
\sum_{j=1}^K \alpha_j f_j(x)
\label{eq:mix_1}
\end{equation}

where the PDFs $f_j$ satisfy a set of constraints associated with a histogram regularization procedure as outlined in section \ref{MC}. Subpopulation fractions $\alpha_j$ (``mixture weights") are required to sum to unity, i.e. $\sum_{j=1}^K \alpha_j=1$.

%
%
%
%

The variable $x$ can correspond to particle pseudorapidity $\eta$, a kinematic variable related to the particle polar angle $\theta$ in the laboratory frame by the expression 
$\eta=-\mbox{ln}(\mbox{tan}\theta/2)$, or $p_T$ i.e. 
particle transverse momentum 
with respect to the beam direction. The subpopulation PDFs $f_j$ are defined in terms of regularized histograms of $x$, as described in section \ref{MC}, where the associated constraints imposed on the PDFs are detailed. The symbol $\varphi_j$ will be used to denote the estimate of the generic subpopulation PDF $f_j$ throughout the text.

The choice of (\ref{eq:mix_1}) was driven by our previous studies, where assuming a predefined PDF functional form led to significant bias on the mixture weight estimates. That bias ultimately related to assuming that PDFs obtained from high-statistics control samples were also appropriate to describe the corresponding probability distributions in a lower-statistics data set. However, fluctuations are sometimes appreciable, and for this reason it is necessary for the model to provide more flexibility if the effect of fluctuations in a data set of interest is to be described. While a rigorous treatment may call for the use of nonparametric Bayesian methods \cite{Bayes_np}, which can provide an additional dimension of flexibility to statistical models, it was decided to adopt a simplified intuition-driven approach for this study, in order to avoid introducing additional complications not related to the algorithm itself in this phase of the development.

The histogram regularization procedure described in section \ref{MC} can be seen as a simplified version of established methods such as Tikhonov regularization \cite{tik}, which can be used to impose smoothness constraints on a likelihood maximization problem. From a conceptual point of view, an alternative way of interpreting the model used in this study is as a simplified version of established kernel or wavelet-based techniques, where regularized histograms effectively play the role of a set of basis functions. In the absence of any constraints to the PDFs in the mixture, the statistical model (\ref{eq:mix_1}) would 
not be well defined, so this is an essential ingredient. Additional remarks about the existence and uniqueness of the stationary distribution for the Markov Chain associated with the algorithm in the configuration used for this study will be provided in section \ref{MC} after the discussion of the Monte Carlo analysis.

Given the mixture of probability distributions (\ref{eq:mix_1}) and a set of observations $\{x_i\}_{i=1,...,N}$, the problem of clustering the latter into $K$ groups by probabilistically associating each of them with a distribution of origin has been solved numerically in a Bayesian framework using MCMC techniques. In particular, the Gibbs sampler \cite{geman}, which directly inspired this work, has been used for this purpose in different disciplines.

The basic pseudocode of the proposed algorithm is reported below. The value of variable $v$ at iteration $t$ is indicated with $v^{(t)}$ throughout.

\begin{enumerate}
\item {\bf Initialization:} Choose $\underline{\alpha}^{(0)}=\{\alpha_j^{(0)}\}_j$ and
$f_j^{(0)}$=$\varphi_j^{(0)}$, $j=1,...,K$ as described in section \ref{MC}.
\item {\bf Iteration $t$:}
\begin{enumerate}
\item Generate the ``allocation variables" 
$z_{ij}^{(t)}$, $i=1,...,N$, $j=1,...,K$ based on probabilities
$P(z_{ij}^{(t)}=1 | \alpha_j^{(t-1)}, \varphi_j^{(t-1)}, x_i)$ proportional to
$\alpha_j^{(t-1)} f(x_i | \varphi_j^{(t-1)})$.
The quantity $z_{ij}^{(t)}$ equals 1 when observation $i$ is mapped to distribution $j$ at iteration $t$, and 0 otherwise. In general, the variables $z_{ij}^{(t)}$ depend both on the mixture weights $\alpha_j$ and on the estimates $\varphi_j$ of the subpopulation PDFs from the previous iteration.
\item Generate $\underline{\alpha}^{(t)}$ from the probability density function of $\underline{\alpha}$ given $\underline{z}^{(t-1)}=\{z_{ij}^{(t-1)}\}_{ij}$, $\rho(\underline{\alpha} | \underline{z}^{(t-1)})$. Knowledge of which particles are mapped to process $j$ at iteration $t-1$ makes it possible to generate the subpopulation fractions $\underline{\alpha}$ at iteration $t$.
\item Obtain an updated estimate of the subpopulation PDFs from the data $\underline{x}$ based on the knowledge of which particles are mapped to subpopulation $j$ at iteration $t-1$. Details are provided in section
\ref{MC}.
\end{enumerate}
\end{enumerate}
A specific choice for the function $\rho$ and a way to obtain updated estimates of the subpopulation PDFs $f_j$ are described in section \ref{MC} with reference to the Monte Carlo study.

The central idea of the algorithm is the following: the better the observations $\{x_i\}_i$ are mapped to the subpopulations $j=1,...,K$, the more accurate the estimates of $\rho(\underline{\alpha} | \underline{z})$ and of the subpopulation PDFs 
$f_j$. 
Once some correct values of $z_{ij}$ are found, $\rho(\underline{\alpha} | \underline{z})$ and $\varphi_j$ begin to roughly reflect the correct distributions, which in turn leads to additional correct mappings $z_{ij}$ to be found at subsequent iterations.


The above pseudocode is very similar to the Gibbs sampler, where updated estimates of subpopulation PDFs are obtained at each iteration, as indicated at step (c). On the other hand, when step (c) is removed from the pseudocode, particles are mapped to signal and background based on the subpopulation PDFs provided at initialization, and the algorithm is then more akin to EM. Throughout the paper we will refer to the two versions of the algorithm with step (c) included or not in the pseudocode as ``unconstrained sampler" and ``constrained sampler", respectively.

The primary objective of this article is to study the use of the proposed sampling technique in different configurations in order to:

\renewcommand{\theenumi}{\roman{enumi}}
\begin{enumerate}
\item Obtain estimates $\varphi_j$ of the subpopulation PDFs from the input data set.
\item Estimate the subpopulation weights $\alpha_j$. In the context of this study, this corresponds to estimating the fractions of background and signal particles contained in the input data set.
\item Assign individual particles a probability for them to originate from a given process based on the subpopulation PDFs estimated at step (i) as opposed to relying exclusively on high-statistics templates. In the context of this study, this allows classification of individual particles into signal and background.
\end{enumerate}




\section{Monte Carlo study}
\label{MC}

The algorithm was run on a Monte Carlo data set generated using Pythia 8.140 \cite{pythia1} \cite{pythia2}, obtained superimposing $gg\rightarrow t\bar{t}$ signal events from pp interactions at $\sqrt{s}=14~\mbox{TeV}$ with low-energy strong interactions, so called Minimum Bias events, in order to simulate background. The signal process was chosen to illustrate the use of the algorithm for background discrimination at the particle level. Further studies will be needed in order to extend these results beyond the initial investigation presented in this article, and to assess the potential of population-based techniques for background discrimination in the context of specific analyses at the LHC.

\begin{figure*}
\centering
\subfigure[]{
\includegraphics[scale=0.43]{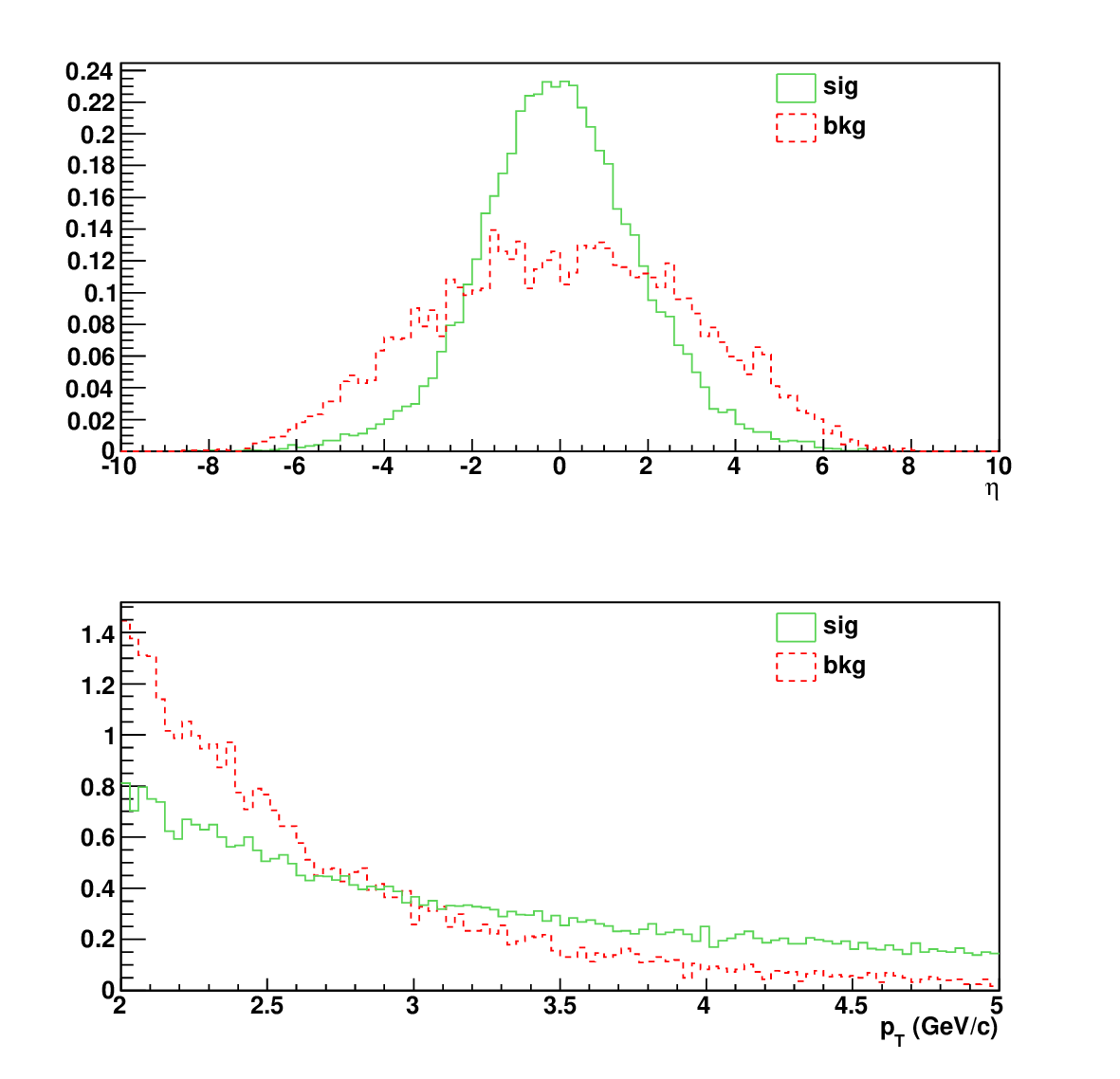}
\label{fig:tt_eta_pt_a}
}
\subfigure[]{
\includegraphics[scale=0.38]{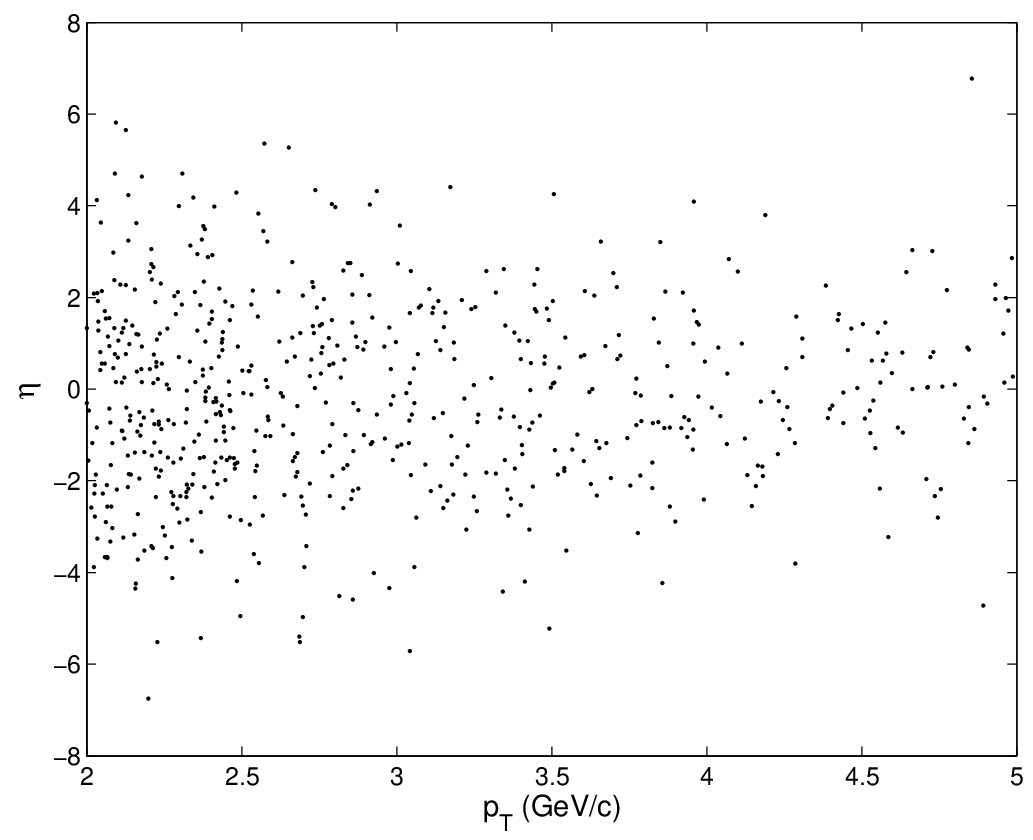}
\label{fig:tt_eta_pt_b}
}
\caption[]{(a) Generator-level $\eta$ and $p_T$ distributions for signal (solid green histograms) and background particles (dashed red histograms) with $2~\mbox{GeV/c}<p_T<5~\mbox{GeV/c}$ from the high-statistics control sample. The distributions correspond to a total number of $\sim 33,000$ particles and are normalized to unit area. (b) The corresponding two-dimensional distribution.}
\label{fig:tt_eta_pt}
\end{figure*}

\begin{figure*}
\centering
\includegraphics[scale=0.38]{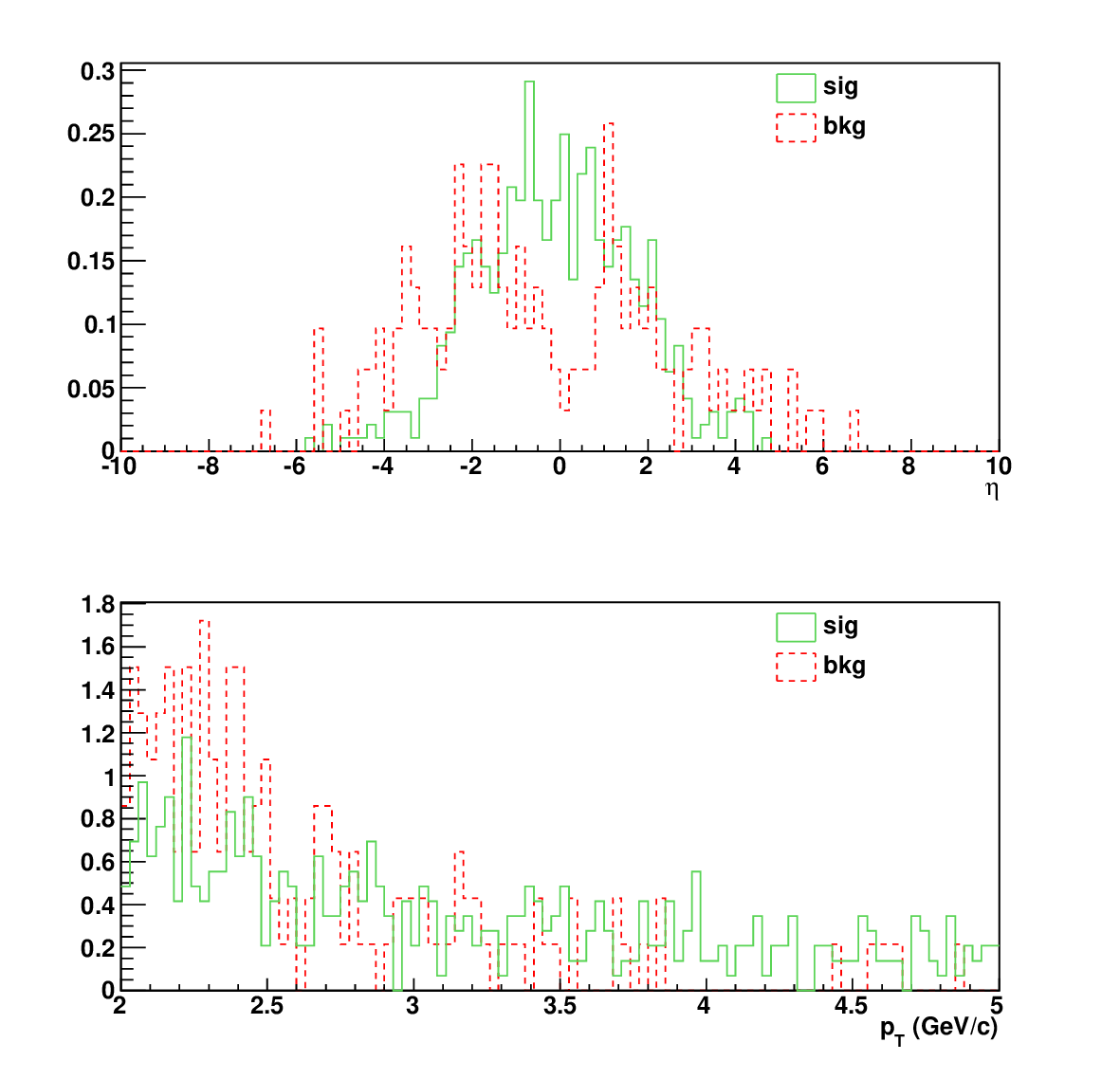}
\includegraphics[scale=0.33]{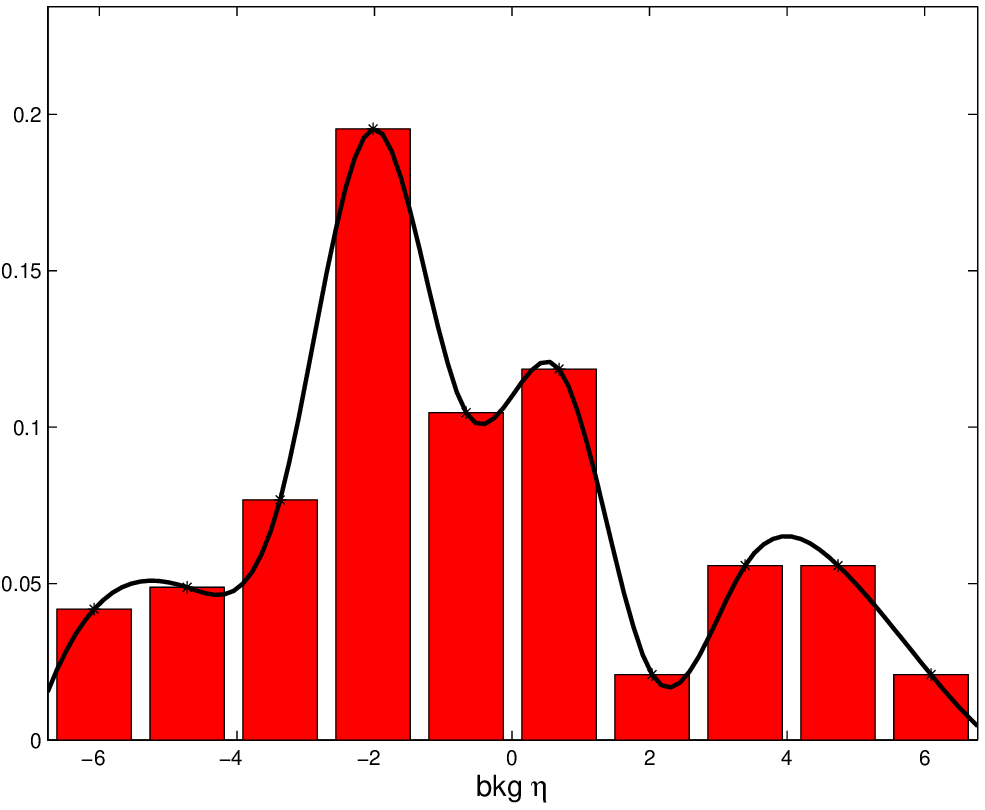}
\caption[]{(a) Particle $\eta$ and $p_T$ distributions from the Monte Carlo input data set used in this study. Solid green and dashed red histograms correspond to signal and background, respectively. Distributions are normalized to unit area. (b) Example of the pseudorapidity $\eta$ distribution of particles mapped to the background subpopulation at a given iteration of the algorithm. The superimposed curve is the result of the regularization procedure described in the text, and is used by the algorithm as an estimate of the corresponding subpopulation PDF.}
\label{fig:myPU_ACAT}
\end{figure*}

The sampler was run over a collection of charged particles with $2~\mbox{GeV/c}<p_T<5~\mbox{GeV/c}$, and individual particles were assigned a probability for them to originate from signal as opposed to background based on their $\eta$ and $p_T$ values. 

The pseudocode of the algorithm used for this application is shown below. Subscripts $sig$ and $bkg$ relate to signal and background, respectively.

\begin{enumerate}
\item {\bf Initialization:} Set $\alpha_{bkg}=\alpha^{(0)}_{bkg}=0.5$, $f_j=\varphi^{(0)}_j$, $j=1,2$. Initial conditions for the estimates $\varphi^{(0)}_j$ of the subpopulation PDFs 
$f_j$ 
are given by regularized $\eta$ and $p_T$ distributions obtained from the high-statistics control sample, as described in section \ref{regular}.
\item {\bf Iteration $t$:} 
\begin{enumerate}
\item Generate $z_{ij}^{(t)}$ for all particles ($i=1,...,N$) and distributions ($j=1,2$ corresponding to background and signal, respectively) according to $P(z_{ij}^{(t)}=1 | \alpha_j^{(t-1)}, \varphi_j^{(t-1)},x_i)\propto \alpha_j^{(t-1)}f_j(x_i | \varphi_j^{(t-1)})$, where $\alpha_1=\alpha_{bkg}$, $\alpha_2=1-\alpha_{bkg}$. 
\item Set $\alpha_j^{(t)}=\sum_i z_{ij}^{(t-1)}/N$, $j=1,2$. This corresponds to the simplest choice of setting $\rho(\alpha_j | \underline{z}^{(t-1)}) = \delta(\alpha_j-\sum_i z_{ij}^{(t-1)}/N)$ for the probability density function of $\underline{\alpha}$ given $\underline{z}$.
\item Obtain updated estimates of the subpopulation PDFs by regularizing the $\eta$ and $p_T$ distributions corresponding to particles mapped to the relevant subpopulation at iteration $t-1$, i.e. based on $z_{ij}^{(t-1)}$.
\end{enumerate}
\end{enumerate}

In general, the functions $f_j$ are the joint PDFs for $\eta$ and $p_T$ corresponding to background ($j=1$) and signal particles ($j=2$). This study is restricted to charged particles with $2~\mbox{GeV/c}<p_T<5~\mbox{GeV/c}$, which makes it possible to neglect the correlation between $\eta$ and $p_T$ as a first approximation. For this reason, the joint PDFs take the form $f_{sig/bkg}=f^{(\eta)}_{sig/bkg}f^{(p_T)}_{sig/bkg}$, and obtaining updated estimates of the subpopulation PDFs reduces to regularization of one-dimensional histograms, as described in the following.

As for the number of iterations to be used with the algorithm, no rule is documented in the statistics literature with reference to related techniques, and the choice is generally problem-dependent. The number of iterations was set to 1,000 in this study, and probabilities were averaged over the last 100. Runs were also performed letting the sampler run for a longer time: the algorithm exhibited a relatively-fast convergence on the data set analyzed, and no gain was found by choosing a higher number of iterations. Moreover, multiple runs were performed corresponding to different initial conditions in order to make sure that the algorithm always converged. In particular, the initial conditions for the subpopulation PDFs were perturbed by using different initial conditions for the fits to the high-statistics distributions from the control sample. Similarly, the generation parameters in the toy Monte Carlo study were varied around their nominal values by $\pm 10\%$, with no appreciable difference in the results.


In order to obtain initial conditions $\varphi_j^{(0)}$ for the subpopulation PDFs $f_j$, a Monte Carlo data set was used containing a total of about 33,000 charged particles in the kinematic range $2~\mbox{GeV/c}<p_T<5~\mbox{GeV/c}$. In addition to estimation of $\varphi_j^{(0)}$, this high-statistics control sample was also used to guide the histogram regularization procedure as described in the following. Figure \ref{fig:tt_eta_pt} (a) shows the $\eta$ and $p_T$ distributions for signal and background particles from the control sample (solid green and dashed red histograms, respectively).

As anticipated, one of the goals of the sampler is to estimate the shapes of the signal and background PDFs from the input collection of particles. This way, the algorithm will be able to classify particles into signal and background without relying exclusively on predefined templates: the background PDFs estimated by the algorithm are expected to reflect the specific background conditions in the input data set, which can sometimes be notably different from the average conditions of a high-statistics control sample, where the effects of fluctuations are typically washed out.


The algorithm basically uncovers a signal and a background subpopulation in the input collection of particles based on the data and on initial conditions on the subpopulation PDFs. The results presented in this article relate to an input data set comprising 636 charged particles in the kinematic region $2~\mbox{GeV/c}<p_T<5~\mbox{GeV/c}$, out of which 481 originate from a signal hard process and 155 from Minimum Bias, corresponding to a fraction of background particles of $\sim24\%$. The total number of particles in the input data set is in line with typical charged particle multiplicities at the LHC as of July 2011, when this analysis was performed.

The signal and background $\eta$ and $p_T$ distributions corresponding to the Monte Carlo input data set used in this study are shown in figure \ref{fig:myPU_ACAT} (a). The solid green (dashed red) histograms in the upper panel display the signal (background) $\eta$ distributions, normalized to unit area. The corresponding $p_T$ distributions are given in the lower panel. It is worth noticing that some of these distributions are appreciably different from the corresponding ones obtained from the control sample due to the presence of fluctuations in the data, as expected. In particular, the background $\eta$ distribution exhibits two modes that are shifted with respect to zero, while the corresponding distribution from the control sample is centered around zero.


In order to illustrate the histogram regularization procedure used in this study, figure \ref{fig:myPU_ACAT} (b) shows an example of the $\eta$ distribution of particles mapped to the 
background 
subpopulation at a given iteration of the algorithm. As opposed to assuming a functional form for the PDF and fitting a function to the histogram, the histogram is regularized, i.e. the subpopulation PDF is obtained by means of spline interpolation of the histogram contents as further discussed in the following. The superimposed curve in the figure corresponds to the regularized histogram, and is used by the algorithm as an estimate of the corresponding subpopulation PDF.


As anticipated, this approach gives the algorithm more flexibility when estimating the shapes of the subpopulation PDFs from the input data set with respect to our previous attempts that relied on a predefined PDF functional form, while still leading to a well-defined target distribution for the associated Markov Chain.

\subsection{Regularization}
\label{regular}

Step (c) in the pseudocode shown in section \ref{MC} requires iterative PDF updates based on the current mapping of individual particles to different subpopulations. This operation is performed when the algorithm is operated in unconstrained mode, as discussed 
above. 


As anticipated, it was decided to adopt a simplified statistical model for the purpose of this investigation, while at the same time providing enough flexibility for the algorithm to be able to describe fluctuations. In the context of this study, this was done by performing spline interpolation of one-dimensional $\eta$ and $p_T$ histograms. As previously mentioned, this can be seen as a simplified version of established regularization techniques, for instance as a way to use a priori information about the underlying distributions in order to get rid of spurious oscillatory components. Such methods have been analyzed in detail in particle physics in order to develop unfolding procedures, with a view to ``removing" detector effects from observed distributions, see e.g. \cite{regul}.

The complexity of the histogram regularization procedure used in this study was intentionally kept minimal in order to avoid the introduction of additional complications that might obscure the response of the algorithm at this stage of the development. Further studies will be needed in order to understand in detail how the 
results are affected by the regularization procedure.

\begin{figure*}
\centering
\includegraphics[scale=0.33]{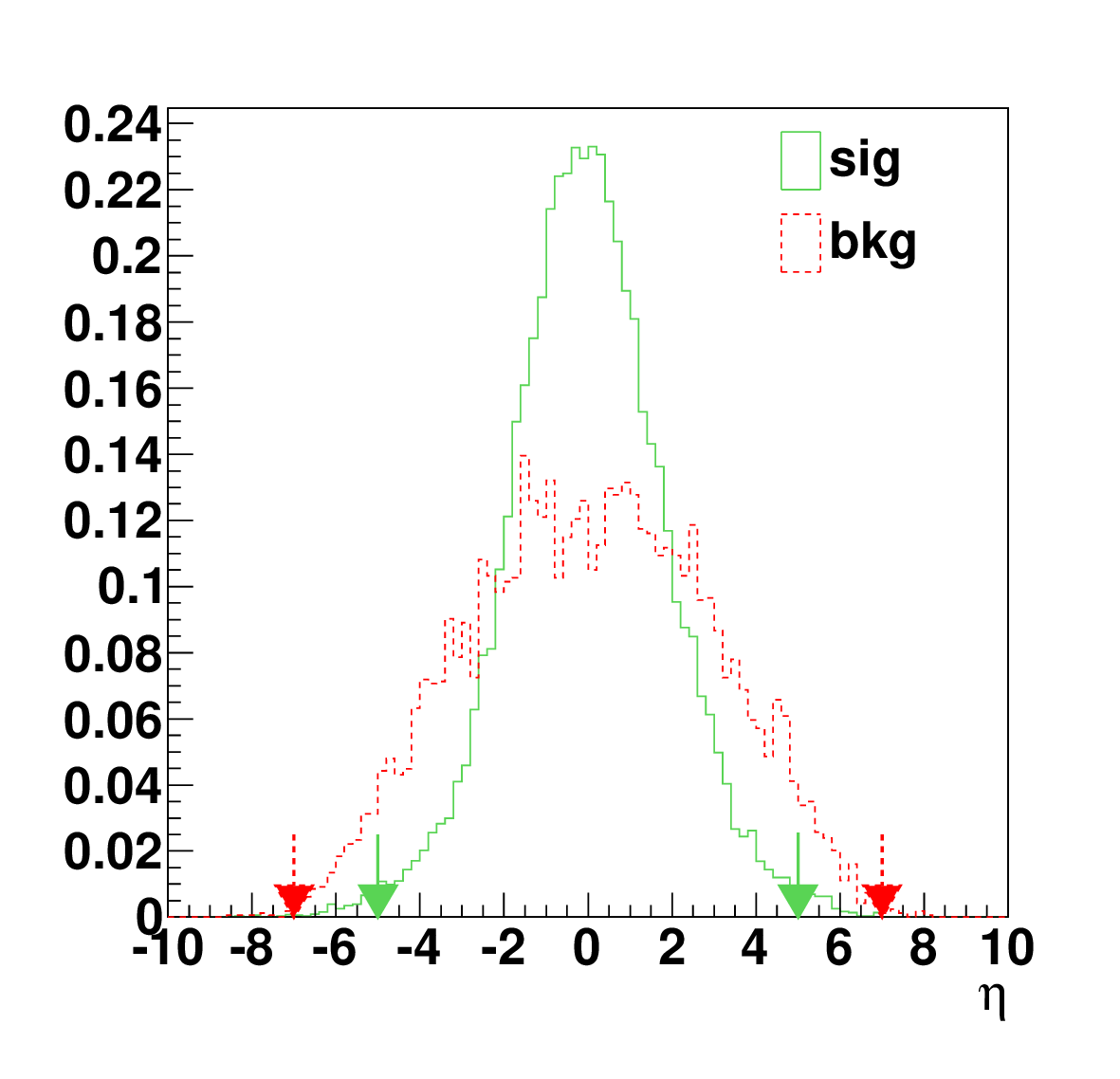}\hspace{3pc}%
\begin{minipage}[b]{16pc}\caption{\label{fig:regul_w}Illustration of the regularization window used in this study. The histograms correspond to the signal (solid green) and background (dashed red) $\eta$ distributions from the high-statistics control sample. The positions of the solid green and dashed red arrows correspond to the regularization window: at each iteration of the algorithm, only the part of the distribution that lies between the arrows is used for spline interpolation, which makes the results robust against outliers. Additional details are given in the text.}
\end{minipage}
\end{figure*}

In the context of this investigation, a priori information about the signal and background PDFs was obtained from the high-statistics control sample. When the subpopulation PDFs are updated iteratively during the execution of the sampler, i.e. when the algorithm is operated in unconstrained mode, a ``regularization window" is applied to the $\eta$ histograms in order to get rid of outliers: in other words, a spline interpolation of the histogram contents is obtained using only the part of the histogram that lies between a minimum and a maximum $\eta$ value, which leads to extreme fluctuations on the tails of the distribution to be excluded. It is worth noticing that assuming a functional form for the PDF and fitting it to the histogram would effectively produce a similar result, i.e. it would reduce the impact of outliers on the estimated PDF. However, as anticipated, that approach was observed to introduce significant bias in previous studies, and was thus abandoned in favor of the statistical model presented in (\ref{eq:mix_1}), where subpopulation PDFs are defined as the output of a histogram regularization procedure without reference to any predefined functional form, but still subject to regularization constraints. Figure \ref{fig:regul_w} shows signal (solid green) and background (dashed red) $\eta$ distributions from the high-statistics control sample, with superimposed arrows indicating the regularization window. The maximum $\left|\eta\right|$ value was set to $\left|\eta\right|=5$ ($\left|\eta\right|=7$) for signal (background).


On top of this, again with a view to getting rid of extreme fluctuations when regularizing histograms, boundary conditions were introduced on the $\eta$ and $p_T$ PDFs, constraining the value of $f_j$ to points chosen based on the 
control sample distributions: the signal (background) $\eta$ PDF was constrained to 0 when $\left|\eta\right|>5$ ($\left|\eta\right|>7$), and signal (background) $p_T$ PDFs were constrained at $2~\mbox{GeV/c}$ and $5~\mbox{GeV/c}$ to 0.7 (1.2) and 0.1 (0) (see figure \ref{fig:tt_eta_pt}(a)).

The 
results were found to be stable with respect to reasonable changes to the above regularization constraints.

\subsection{Choice of configuration}
\label{GS:config}

As anticipated, the algorithm can be operated in unconstrained or in constrained mode, depending on whether step (c) in the pseudocode given in section \ref{MC} is included or not.

As already pointed out in section \ref{GS}, the algorithm 
processes an input collection of particles in order to obtain one or more of the following results:

\renewcommand{\theenumi}{\roman{enumi}}
\begin{enumerate}
\item Estimate the shapes of the subpopulation PDFs from the input data set.
\item Estimate the fraction of particles associated with a given process in the input data set, e.g. the fraction of background particles.
\item Assign individual particles a probability for them to originate from a given process, such as a hard scattering of interest as opposed to background 
associated with low-energy strong interactions, based on the subpopulation PDFs estimated at step (i) as opposed to relying on predefined templates that only reflect average background conditions.
\end{enumerate}

Depending on the objective, it may be appropriate to run the algorithm in different modes.

For instance, the histogram regularization procedure that is used here to obtain iterative estimates of the subpopulation PDFs when the algorithm is operated in unconstrained mode inherently leads to a 
bias on the 
mixture weights, because imposing a regularization window changes the number of particles that are mapped to signal or background at a given iteration. For this reason, it may be more appropriate to use a different approach to estimate the fraction of background particles.

One option is described below:

\begin{itemize}
\item[(a)] The constrained sampler is first used to estimate the mixture weights. In the two-subpopulation scenario described in this study, goal (ii) above corresponds to estimating the fraction of background particles contained in the input data set. The initial conditions for the mixture weights are $\alpha_1^{(0)}=\alpha_2^{(0)}=0.5$, corresponding to no prior knowledge about the fraction of background particles in the input sample. The subpopulation PDFs are kept fixed at the estimates provided by the high-statistics control sample. The corresponding results are described in section \ref{sGS}.
\item[(b)] The algorithm is then run again on the input data set in unconstrained mode, i.e. subpopulation PDFs are now updated at each iteration, starting from initial conditions corresponding to regularized distributions from the high-statistics control sample. However, the 
mixture weights are now kept fixed at the results from the previous step.
\end{itemize}

It is worth noticing that the algorithm differs from a proper Gibbs sampler in both cases.

As for assigning individual particles in the input data set a probability for them to originate from signal as opposed to background, the most appropriate approach may again depend on the specific application. In general, probabilities may be assigned directly using the unconstrained sampler at step (b) above, as done in this study, or an additional run of the algorithm in constrained mode may alternatively be added after the previous two, with fixed PDFs given by the estimates from step (b). Further studies will be necessary in order to better understand the classification performance of the algorithm in different configurations and to guide this choice.

The 
results obtained running the constrained and the 
unconstrained sampler as described above on the Monte Carlo input data set used in this study are reported and discussed in the following sections.


\subsubsection{Constrained sampler}
\label{sGS}

\begin{figure*}
\centering
\begin{minipage}{16pc}
\includegraphics[scale=0.38]{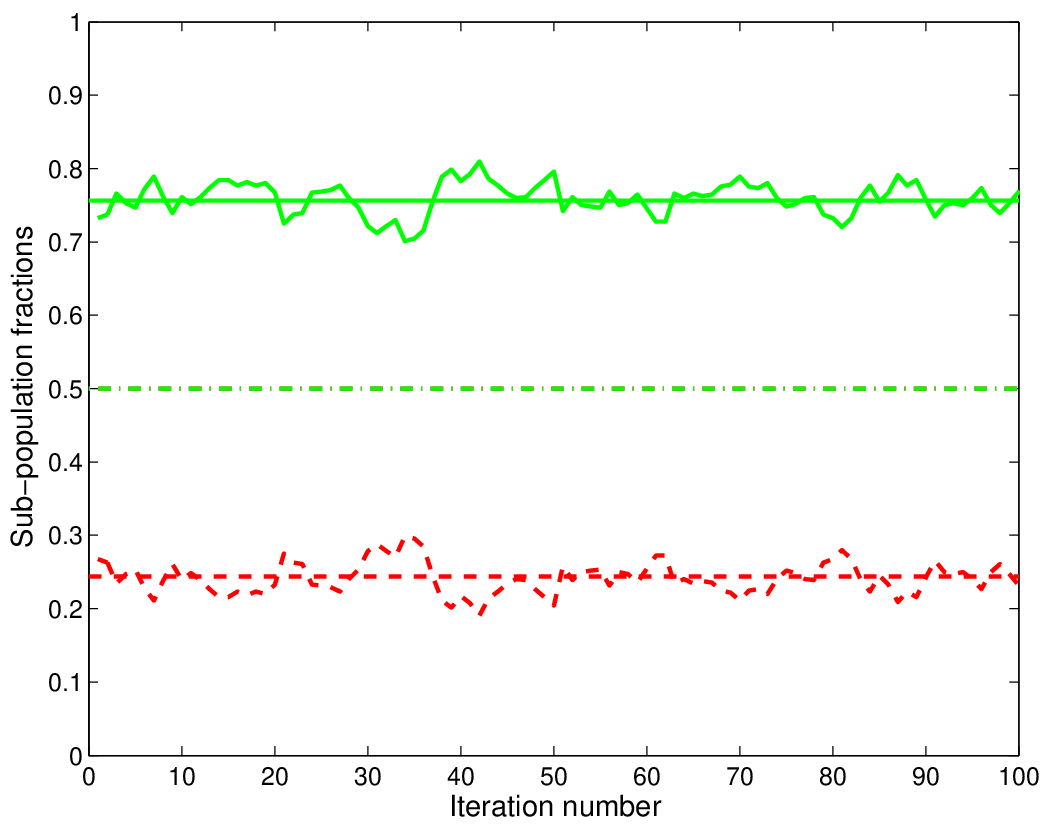}
\caption{\label{fig:weights_s}Mixture weights obtained running the constrained sampler on the Monte Carlo input data set. Results from the last 100 iterations are shown. The solid green (dashed red) curve denotes the estimated fraction of signal (background) particles. The solid green (dashed red) horizontal line indicates the true value for signal (background) from the simulation, and the dash-dot line corresponds to the initial conditions for the mixture weights.}
\end{minipage}\hspace{3pc}%
\begin{minipage}{16pc}
\includegraphics[scale=0.38]{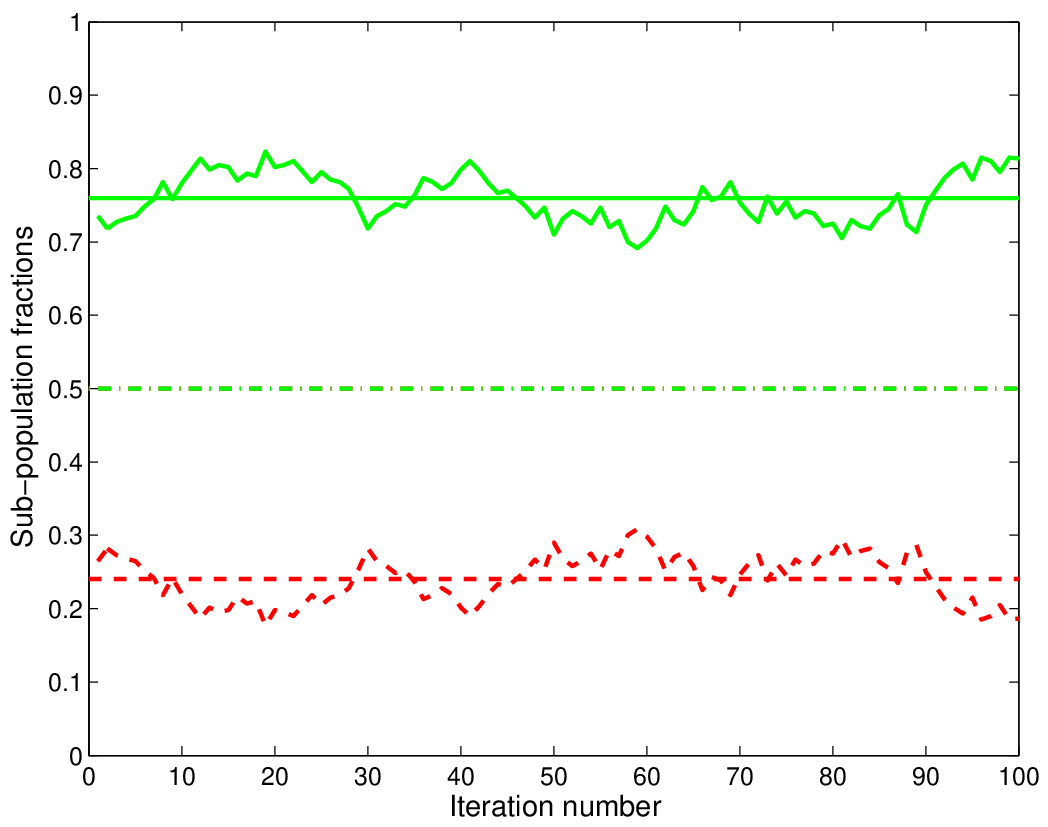}
\caption{\label{fig:weights_toy}Mixture weights obtained running the constrained sampler on a toy Monte Carlo data set, as described in the text. Results from the last 100 iterations are shown. The solid green (dashed red) curve corresponds to the estimated fraction of signal (background) particles. The solid green (dashed red) horizontal line indicates the true value for signal (background) from the toy Monte Carlo, and the dash-dot line corresponds to the initial conditions for the mixture weights.}
\end{minipage}
\end{figure*}

As anticipated, the algorithm in constrained mode was primarily used in this study in order to estimate the mixture weights, i.e. the fraction of background particles in the input data set. Figure \ref{fig:weights_s} shows the corresponding estimates over the last 100 iterations. The solid green and dashed red curves correspond to the estimated fractions of signal and background particles, respectively. The solid green (dashed red) horizontal line indicates the signal (background) true value from the simulation, while the dash-dot line corresponds to the initial conditions for the mixture weights.

Additional runs on toy Monte Carlo samples were performed as a cross-check, as described in the appendix. Figure \ref{fig:weights_toy} displays the estimated mixture weights obtained by running the constrained sampler on a toy Monte Carlo data set with subpopulation PDFs kept fixed at truth information.

\subsubsection{Unconstrained sampler}

The unconstrained sampler was used in this study to estimate the shapes of the signal and background PDFs from the input data set, while keeping the mixture weights fixed at the results obtained from the previous run of the algorithm in constrained mode.

\begin{figure*}
\centering
\subfigure[]{
\includegraphics[scale=0.41]{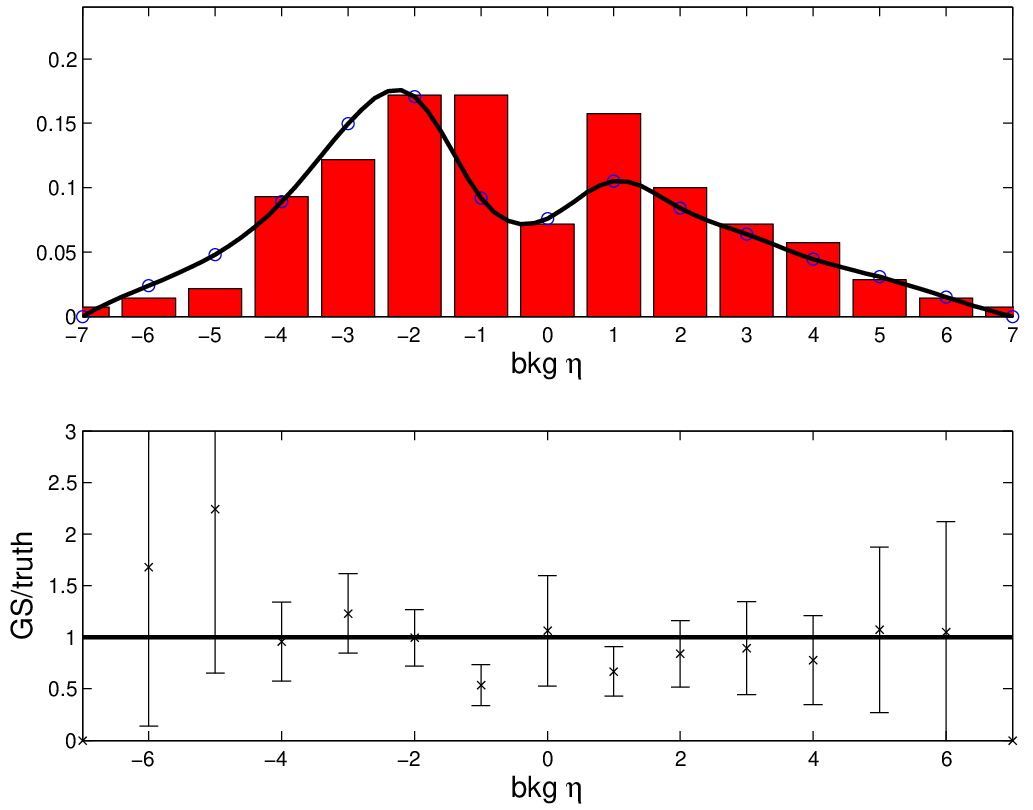}
\label{fig:splines_us_a}
}
\subfigure[]{
\includegraphics[scale=0.41]{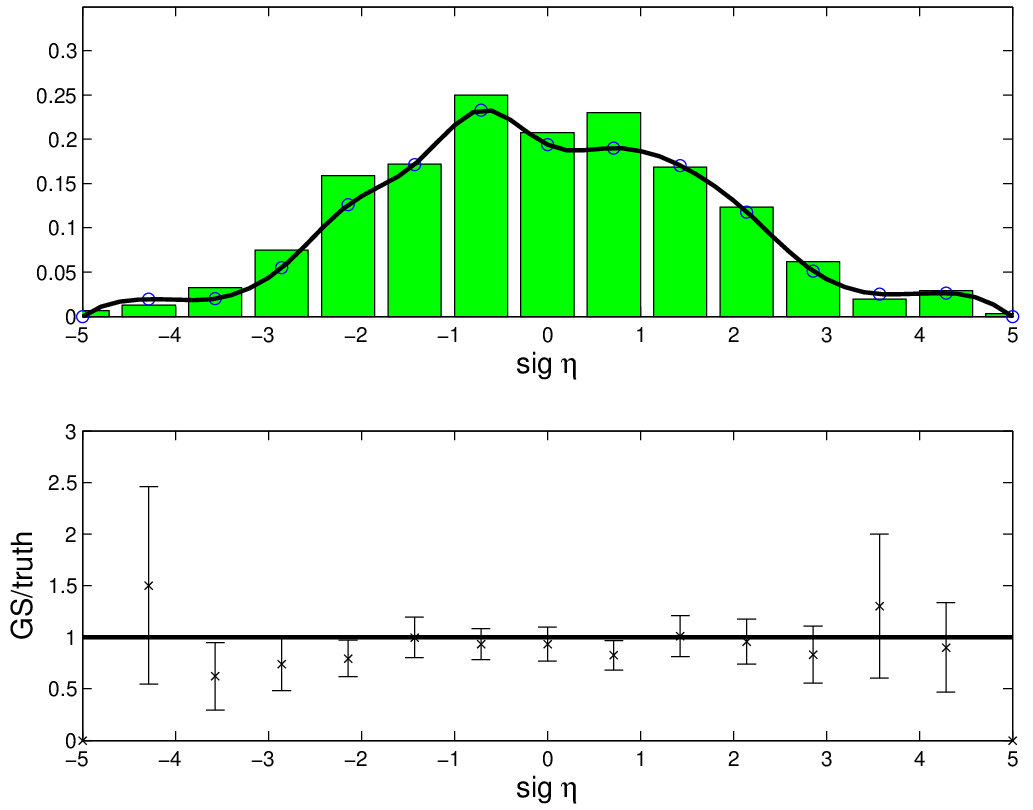}
\label{fig:splines_us_b}
}
\subfigure[]{
\includegraphics[scale=0.41]{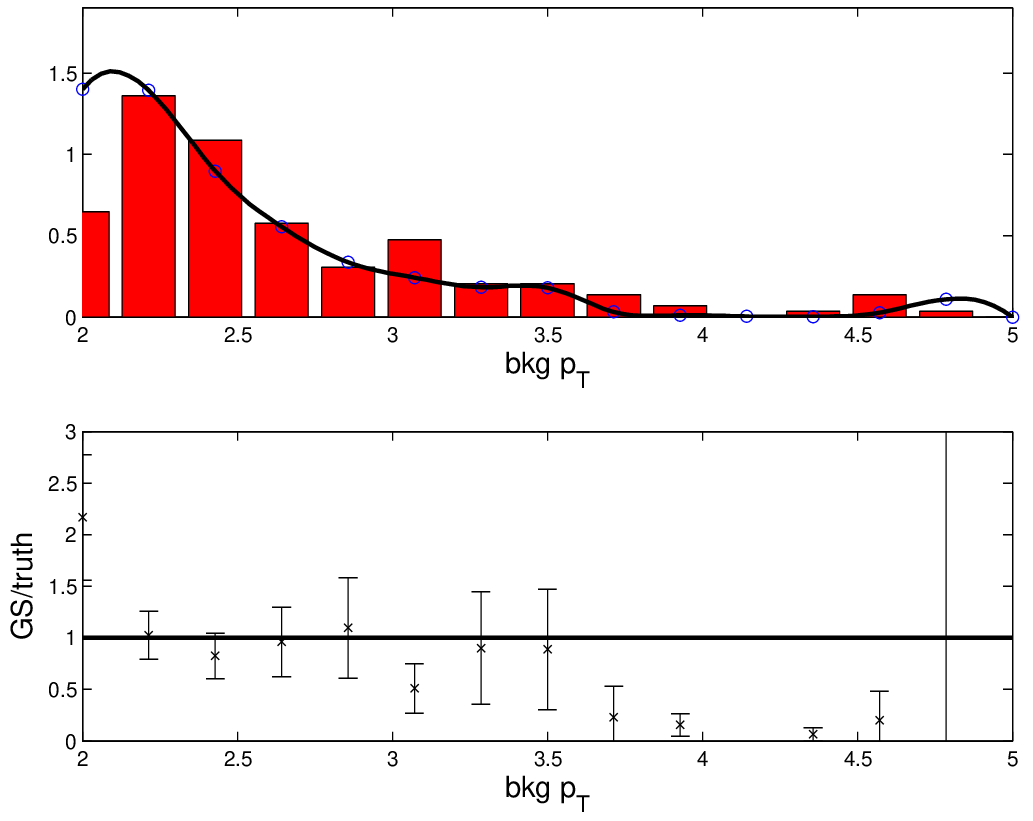}
\label{fig:splines_us_c}
}
\subfigure[]{
\includegraphics[scale=0.41]{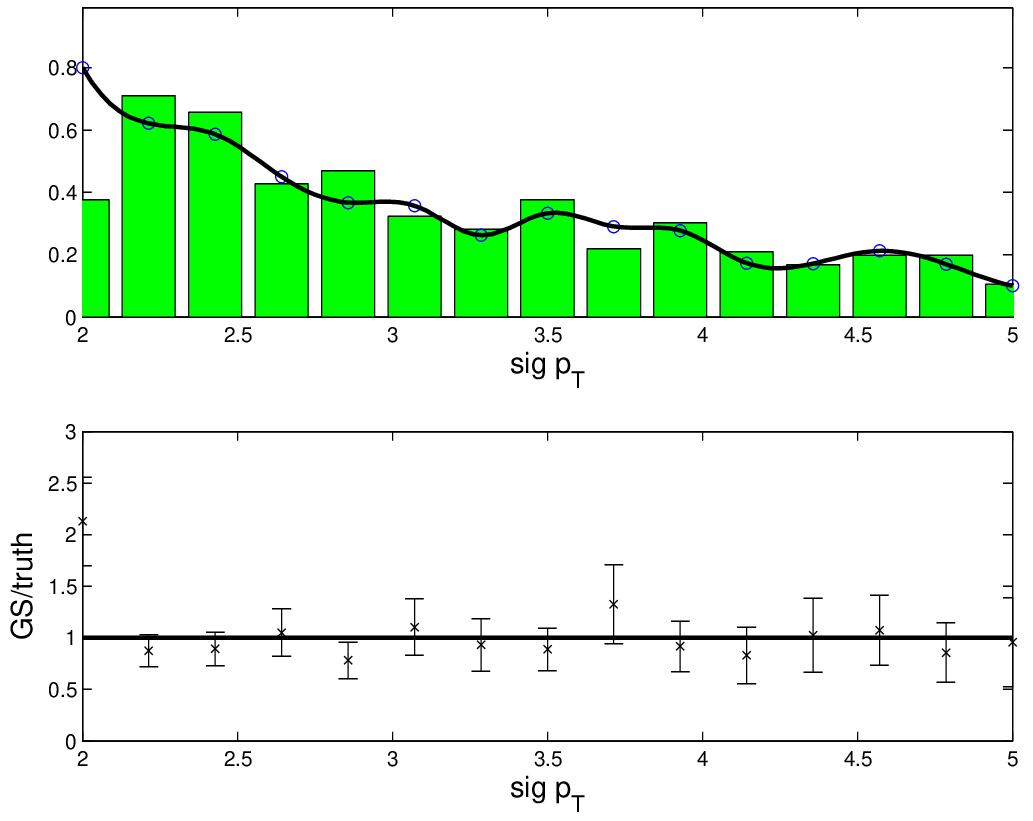}
\label{fig:splines_us_d}
}
\caption[]{Subpopulation PDFs estimated by the unconstrained sampler on the Monte Carlo input data set used in this study, averaged over the last 100 iterations. (a) Background $\eta$. (b) Signal $\eta$. (c) Background $p_T$. (d) Signal $p_T$. In each subfigure, the upper panel shows truth information (histogram bars) superimposed with the result of the regularization procedure averaged over the last 100 iterations (curve). The lower panels display the ratio between the subpopulation PDFs estimated by the algorithm and the corresponding truth-level information.}
\label{fig:splines_us}
\end{figure*}

Figure \ref{fig:splines_us} shows the subpopulation PDFs estimated by the algorithm on the Monte Carlo input data set. The curves correspond to the output of the histogram regularization procedure averaged over the last 100 iterations, superimposed on the true distributions (histograms). The $\eta$ ($p_T$) distributions are displayed in the top (bottom) plots, figures on the left-hand (right-hand) side correponding to background (signal). All distributions are normalized to unit area. The bottom panel in each figure shows the corresponding ratio between the relevant subpopulation PDF estimated by the algorithm and truth information.

The figure illustrates a distinctive characteristic of the proposed algorithm as compared to well-established techniques. As already pointed out, the background $\eta$ distribution in the Monte Carlo data set used in this study differs appreciably from the corresponding distribution obtained from the control sample, as shown by the two modes around $\eta\simeq -2$ and $\eta\simeq 1$ in the figure, as opposed to the symmetric distribution centered around $\eta=0$ that is obtained from the high-statistics data set. As it can be seen, the sampler was able to identify those deviations with respect to the control sample template. Such properties of the background PDF are specific to the data set under investigation, and could not have been estimated using the control sample, since that reflects background conditions averaged over a large number of particles.

\begin{figure*}
\centering
\begin{minipage}{15pc}
\includegraphics[scale=0.38]{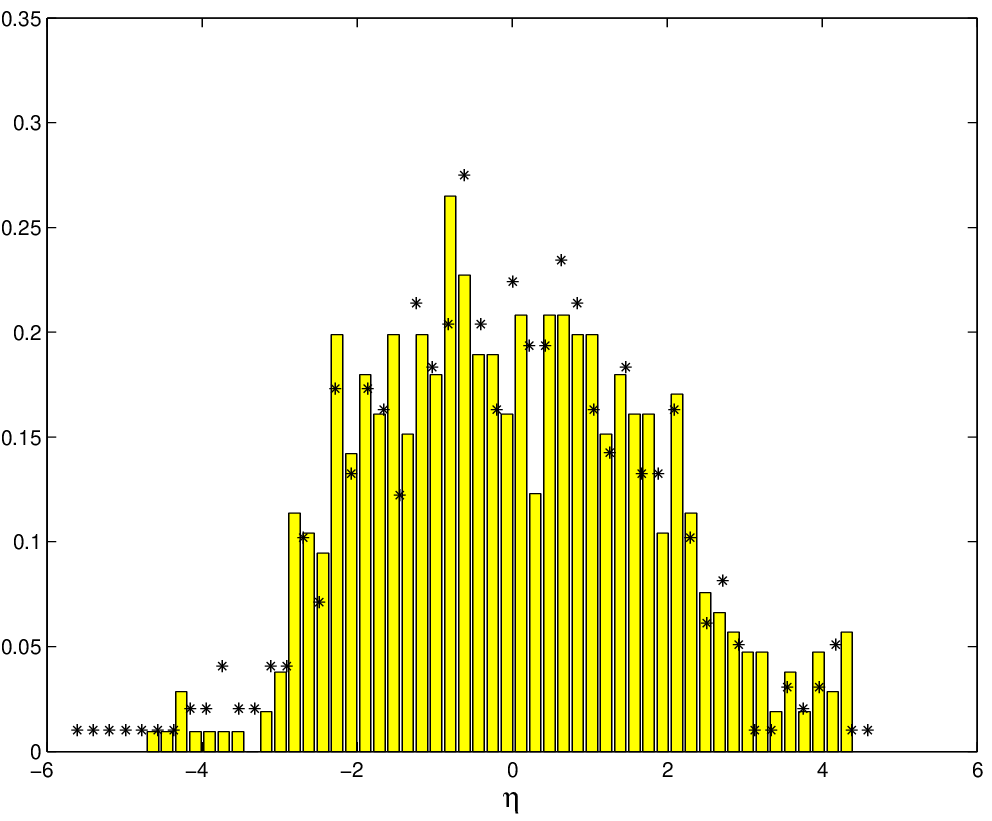}
\caption{\label{fig:tt_valid_2}Comparison between the $\eta$ distribution for particles with $P_{sig}>0.5$ (histogram bars) and the corresponding distribution from truth (stars). Additional information is given in the text.}
\end{minipage}\hspace{2pc}%
\begin{minipage}{18pc}
\includegraphics[scale=0.38]{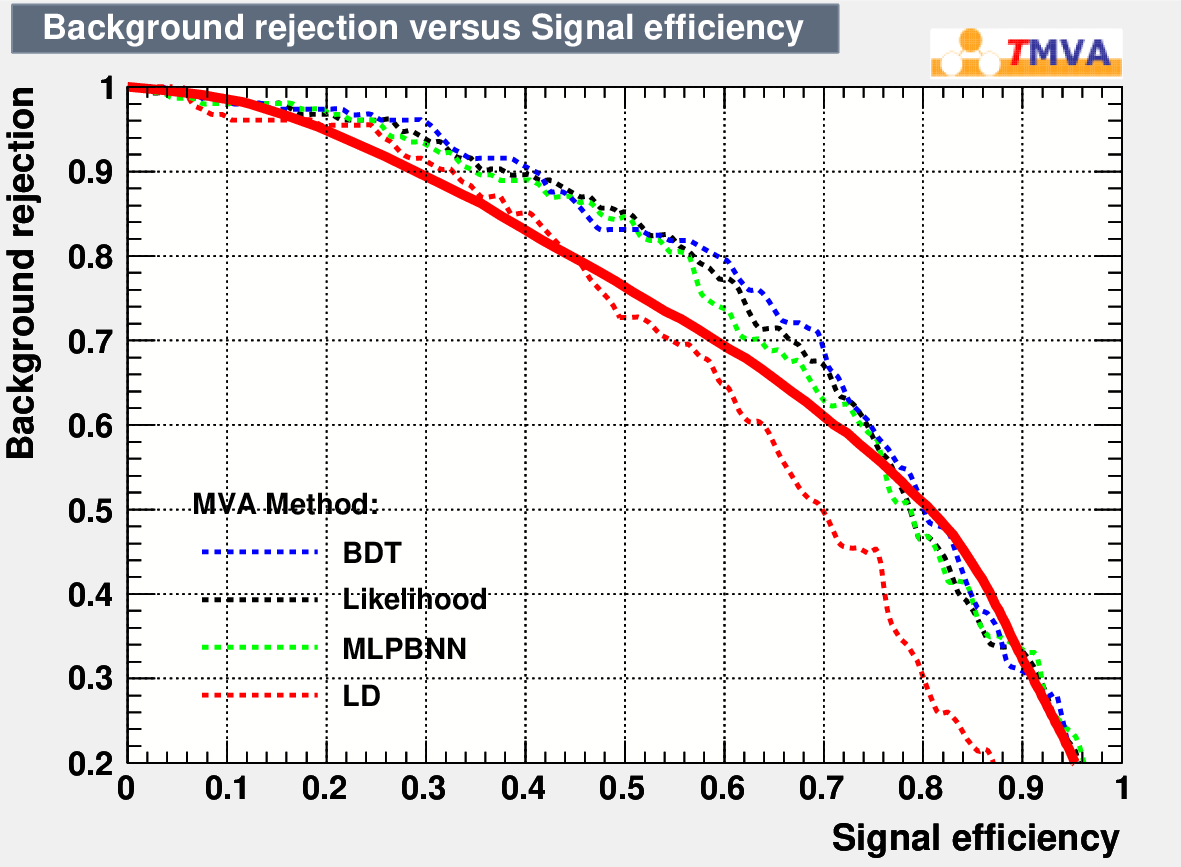}
\caption{\label{fig:tt_ROC_cfr}Comparison between the ROC curve obtained using the unconstrained sampler, shown here as background rejection rate as a function of signal efficiency, and the corresponding curves obtained using existing supervised classification techniques from TMVA, as described in the text. The solid red line is the curve from the unconstrained sampler corresponding to the average over the last 100 iterations. The other curves correspond to TMVA algorithms, namely Boosted Decision Trees (dashed blue), Naive Bayes classification (dashed black), the Neural Network-based classifier MLPBNN (dashed green), and Linear Discriminant (dashed red). Additional information is given in the text.}
\end{minipage}
\end{figure*}

In addition to obtaining data-driven estimates of the subpopulation PDFs in unconstrained mode, one of the goals of the algorithm in this application is to assign individual particles a probability for them to originate from a given process, such as a hard scattering of interest as opposed to Minimum Bias. In this study, those probabilities were obtained from the same unconstrained run of the algorithm that provided the PDF estimates shown in figure \ref{fig:splines_us}. In general, other choices are possible, such as performing an additional run of the algorithm in constrained mode with subpopulation PDFs kept fixed at the estimates shown in figure \ref{fig:splines_us}, as previously mentioned. 

Detailed studies will be necessary in order to understand the implications of different choices before population-based tools for background discrimination can be applied to physics analysis at the LHC. An initial comparison of the classification performance of the algorithm in the configuration chosen for this study with the corresponding performance of existing techniques is described in section \ref{GS:ROC}.

However, it should be emphasized that the primary objective of the proposed approach is not to improve on existing methods in terms of classification performance, but rather to estimate the effect of fluctuations on the shapes of particle-level probability distributions in a data set of interest.

The probabilities returned by the algorithm were validated by comparing the true kinematic distributions with the corresponding ones for particles with $P_{sig}>0.5$, $P_{sig}$ being the estimated probability for a given particle to originate from the signal process, averaged over the last 100 iterations. Results are shown in figure \ref{fig:tt_valid_2}, where histogram bars indicate the $\eta$ distribution for particles with $P_{sig}>0.5$ and stars correspond to the true distribution. 

\begin{center}
\begin{table*}
\parbox{9cm}{\caption{Average number of 
primary vertices per event 
(second column) expected at different LHC instantaneous luminosities (first column) \cite{lockman}. A 25~ns bunch crossing is assumed. The third column reports the corresponding ratios between the number of background and signal particles observed in the kinematic region considered in this study. These estimates were used to generate the curves shown in figure \ref{fig:tt_contam}, as described in the text.}
}
\begin{tabular}{l l l}
\hline
$L (\mbox{cm}^{-2}\mbox{s}^{-1})$ & $\left< n_{PU} \right>$ & $N_{bkg}/N_{sig}$ \\
\hline
$10^{33}$ & 2.3 & 0.1 \\
$10^{34}$ & 23.0 & 0.9 \\ 
$10^{35}$ & 230.0 & 4.4 \\
\hline
\end{tabular}
\label{tab:nPU}
\end{table*}
\end{center}


\begin{figure*}
\centering
\includegraphics[scale=0.4]{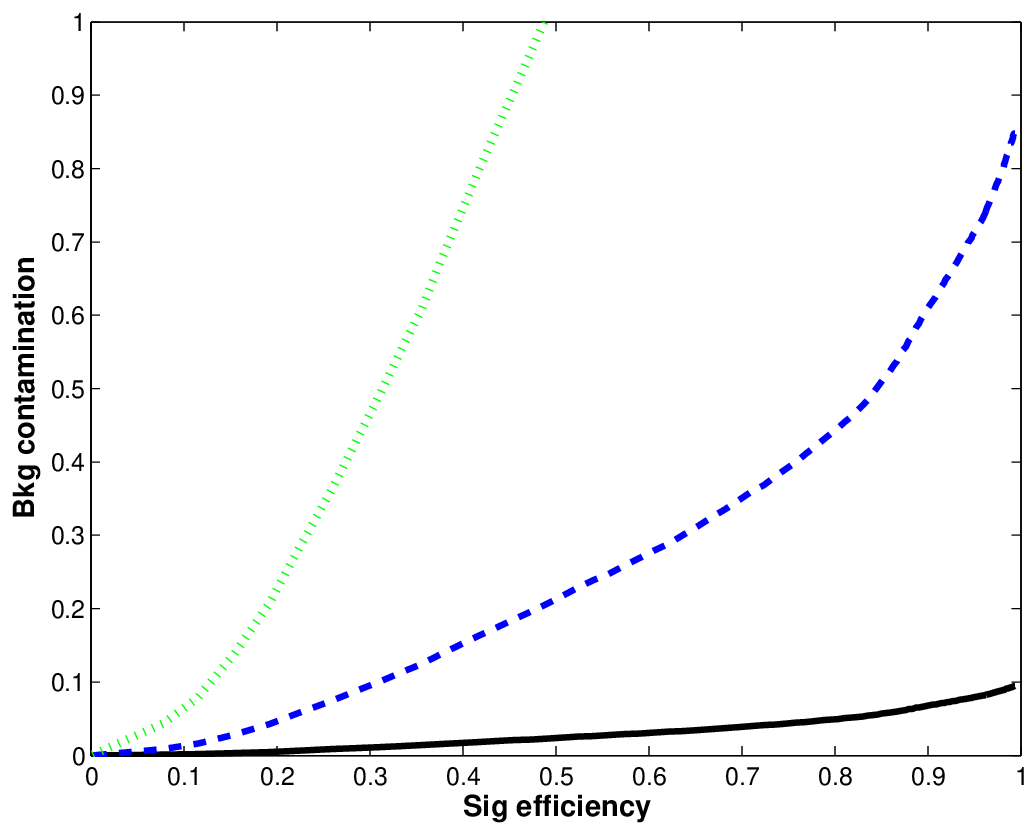}\hspace{2pc}%
\begin{minipage}[b]{16pc}\caption{\label{fig:tt_contam}Background contamination as a function of signal efficiency at different LHC instantaneous luminosities ($10^{33}~\mbox{cm}^{-2}\mbox{s}^{-1}$ solid black, $10^{34}~\mbox{cm}^{-2}\mbox{s}^{-1}$ dashed blue, $10^{35}~\mbox{cm}^{-2}\mbox{s}^{-1}$ dotted green). Background contamination is defined as the 
number of misclassified background particles normalized to the 
number of signal particles. Misclassification probabilities correspond to the ROC curve from the sampler in figure \ref{fig:tt_ROC_cfr}. Additional information is given in the text.}
\end{minipage}
\end{figure*}

\subsection{Classification performance}
\label{GS:ROC}

Operating the sampler as presented in this article is equivalent to using it as a binary classifier. Its performance can thus be quantified using the Receiver Operating Characteristic (ROC) curve, which displays true-positive as a function of false-positive probability. The area under the curve is a number between 0 and 1: the higher its value, the better the classifier is able to discriminate between the two categories (signal and background in this case).
The ROC curve of a random classifier would be a straight line along the main diagonal on the true-positive vs false-positive plane (``chance diagonal" \cite{ROC}).

Figure \ref{fig:tt_ROC_cfr} shows a comparison between the ROC curve obtained using the unconstrained sampler on the Monte Carlo input data set used for this study and the corresponding curves from different supervised multivariate classification methods using TMVA \cite{TMVA} V04-01-00. The curves are displayed using an equivalent representation in terms of background rejection rate as a function of signal efficiency\footnote{Based on our previous terminology, ``background rejection rate" is equivalent to $1-P_{bkg\rightarrow sig}$, and ``signal efficiency" corresponds to $P_{sig\rightarrow sig}$, where $P_{bkg\rightarrow sig}$ ($P_{sig\rightarrow sig}$) is the probability for a background (signal) particle to be mapped to the signal subpopulation.}.

The dashed lines refer to different TMVA methods\footnote{The algorithms were run using the high-statistics control sample for training and the same collection of particles the sampler was run on for testing.}, namely Boosted Decision Trees (dashed blue), Naive Bayes classification (dashed black), the Neural Network-based classifier MLPBNN (dashed green), and Linear Discriminant (dashed red).

The solid red line corresponds to the proposed algorithm. The figure suggests that the classification performance of the sampler is similar to that of existing supervised methods, although other methods perform better in terms of ROC curve on the data set used in this study. However, the advantage of the proposed sampling algorithm with respect to existing methods is not in terms of improved classification performance, but instead relates to estimating features of the signal and background distributions that reflect the presence of fluctuations in the data, which is not possible using established supervised classifiers trained on control samples.

It may also be useful to provide a more precise idea of the background rejections and signal efficiencies that can be achieved using the proposed algorithm corresponding to different LHC instantaneous luminosities\footnote{The expected average number of pile-up interactions, i.e. the expected average number of primary vertices in the events, is here taken as a measure of background activity for illustrative purposes.}. Figure \ref{fig:tt_contam} shows estimates of background contamination as a function of signal efficiency at three different LHC instantaneous luminosities. Background contamination is defined as the number of misclassified background particles normalized to the number of signal particles, and is calculated by rescaling the abscissa of the ROC curve by the ratio between the number of background and signal particles in the kinematic region considered in this study, as given in table 1\footnote{The average numbers of pile-up interactions at different LHC instantaneous luminosities are taken from \cite{lockman}, and correspond to a 25~ns bunch crossing.}. The abscissa of the ROC curve in fact corresponds to false-positive rate, i.e. to the probability for a background particle to be misclassified as signal, and multiplying it by the ratio between the number of background and signal particles provides the desired result. The three curves in figure \ref{fig:tt_contam} correspond to instantaneous luminosities of $10^{33}~\mbox{cm}^{-2}\mbox{s}^{-1}$ (solid black), $10^{34}~\mbox{cm}^{-2}\mbox{s}^{-1}$ (dashed blue), and $10^{35}~\mbox{cm}^{-2}\mbox{s}^{-1}$ (dotted green).



\subsection{Convergence issues}

A remark is necessary with regard to the convergence properties of the Markov Chain associated with the proposed algorithm in the form presented in this article. The proposed technique is here justified primarily based on the results it provides, and based on its ability to estimate the effect of fluctuations on the shapes of particle-level probability distributions in a data set of interest. This is to be compared with the description of background distributions that can be obtained using control samples, which, despite its level of precision, usually only reflects average background conditions and does not take fluctuations into account.


Although the statistical model (\ref{eq:mix_1}) may be questioned from a theoretical point of view and a more rigorous approach based on Bayesian nonparametric methods may be required, the model presented here in practice leads to a well-defined target distribution for the algorithm to sample from. As anticipated, this is primarily due to the constraints associated with the histogram regularization procedure adopted in this study, which effectively restricts the search space and leads to the existence and uniqueness of the stationary distribution of the Markov Chain. This was verified explicitly by using flat distributions as initial conditions for the subpopulation PDFs, making sure that the sampler was still able to estimate the correct PDF shapes.


\subsection{Dependence on the initial conditionus}

One more issue that is worth discussing is the dependence of the results on the initial conditions. The ability to reach the equilibrium distribution regardless of the starting point is a defining feature of Markov Chains. Throughout this study, it has been verified that the initial conditions on the subpopulation PDFs can be perturbed without altering the final resuls. 

It is worth noticing that the similarity of the PDFs estimated by the sampler with the PDF initial conditions obtained from the control sample should not be mistaken for a limitation of the proposed method, but should instead be seen as a defining feature. One of the goals of the algorithm is in fact to improve on the control sample PDF templates. This is done by estimating the effect of fluctuations on the shapes of the probability distributions in the data set under study. For this reason, the PDFs estimated by the sampler are normally similar to the initial PDFs, and the associated Markov Chain generally exhibits a relatively-fast convergence by construction.

\subsection{Concluding remarks}


The possibility to estimate the shapes of signal and background distributions from a data set under study is a distinctive characteristic of the proposed method as compared to existing approaches such as those available in ROOT \cite{ROOT} with TMVA. Although established techniques in some cases provide better classification performance on the data set analyzed in this study, as shown by the comparison in figure \ref{fig:tt_ROC_cfr}, existing methods are in general unable to describe features of the probability distributions that are not already encoded in the training sample. And since the latter typically corresponds to a high-statistics control sample, this usually results in fluctuations in the input data set being neglected.

This technique has been investigated with the prospective goal of developing novel methods for intensive offline analysis of individual events at the LHC, and more generally in particle physics. Data analysis in fact often results in the identification of only a few candidate events that may contain a signal process of interest. 

Traditional methods perform background subtraction based on fixed templates that typically provide a precise description of average background properties. However, this normally neglects features of the probability distributions that are due to fluctuations in the events of interest, and that are normally not present in the control sample templates. We anticipate that the development of dedicated tools for background subtraction based on event-by-event templates thereby taking particle-level fluctuations into account will lead to improved background subtraction and to lower systematic uncertainties. This aspect will be the subject of future studies, as will a quantification of the impact of the algorithm in a realistic analysis environment.

It is also worth noticing that, from a conceptual point of view, the proposed population-based approach is in a sense based on a similar philosophy as Particle Flow, which has been increasingly used in particle physics \cite{partflow}, in that the focus is on individual particles inside events. However, the prospective objective of the proposed technique is totally different, and concentrates on extracting from the data event-by-event particle-level templates that take into account the effect of fluctuations.
%
%
Efforts to eliminate noise in event-by-event analysis of high-energy multiparticle production are reported in the literature, most notably with reference to the study of dynamical fluctuations in heavy-ion collisions, where the notion of ``event-by-event fluctuations" was introduced \cite{noise_QGP_01}, e.g. for mean transverse momentum or mean transverse energy measurement. In the context of such studies, the focus is e.g. on obtaining analytical expressions for the moments of probability distributions that can be used to eliminate statistical fluctuations from the data with a view to extracting information about the underlying dynamics \cite{noise_QGP_02}. Although those studies are conceptually related to the prospective goal of the approach presented in this article in that they aim to subtract noise from individual events, they are fundamentally different. First of all, \cite{noise_QGP_02} requires the 
fluctuations to be Poissonian, while this method works under more general conditions. Moreover, one of the novel aspects of this work is the idea of concentrating on individual particles inside events, reformulating background discrimination in terms of a classification problem at the particle level. The emphasis of this work on a new population-based view of particle physics events is an important aspect that distinguishes the proposed approach from previous efforts.

As a concluding remark, it should also be noted that the iterative nature of the algorithm can lead to a disadvantage with respect to established methods in terms of execution time. However, the running time of the sampler corresponding to 1,000 iterations on the Monte Carlo input data set used in this study was $\sim20~\mbox{s}$ on a 2~GHz Intel Processor with 1~GB RAM, so still reasonable for offline use. In any case, given the parallelization potential of the algorithm, which is a consequence of a similar property of the Gibbs sampler as pointed out in \cite{geman}, improvements may be possible in this respect, for example using commodity Graphics Processing Units (GPUs) that have been used extensively both in particle physics and in other disciplines for compute-intensive applications.

\section{Conclusions and outlook}
\label{concl}

This contribution has presented an initial investigation of a novel approach to background discrimination in particle physics that builds on a population-based view of high-energy particle collision events. Collections of particles are treated as mixtures of subpopulations associated with different physics processes. 
Sampling techniques related to statistical mixture decomposition models are used to assign individual particles a probability for them to originate from a hard scattering of interest as opposed to background 
associated with low-energy strong interactions. This application of the proposed algorithm to a classification problem at the particle level has been pursued with the prospective goal of developing a suite of tools to estimate signal and background properties from individual events at the LHC. 
For instance, a major objective is to obtain estimates of PDF shapes from the data without relying exclusively on templates from high-statistics control samples and without assuming predefined functional forms.

This study has highlighted strengths and limitations of the algorithm operated in different configurations.
In general, systematic uncertainties associated with the use of the algorithm will have to be evaluated in the context of a specific analysis.

Detailed understanding of how the classification performance in different configurations compares to existing techniques will also require further study, as will the possible development of subsequent versions optimized in terms of execution time, building on the inherent parallelizability of the algorithm.

As anticipated, the total number of particles in the Monte Carlo input data set used in this study is in line with typical charged particle multiplicities at the LHC corresponding to operating conditions as of July 2011, when this analysis was performed. For this reason, the results presented in this article are a promising starting point for futher development, with a view to building dedicated software tools for intensive offline analysis of individual events at the LHC.

\section{Acknowledgments}

The author wishes to thank the High Energy Physics Group at the Department of Physics and Astronomy, University College London, and particularly 
Prof. Jonathan M. Butterworth for his precious comments. 
The author also wishes to thank Prof. Trevor Sweeting at the Department of Statistical Science, University College London, for his feedback, and Dr. Alexandros Beskos at the same department for fruitful discussions. Particular gratitude also goes to the Department of Astronomy and Theoretical Physics at Lund University, 
especially 
to Prof. Carsten Peterson and to Prof. Leif Lönnblad for their advice and for fruitful discussions.

\section*{Appendix: Toy Monte Carlo studies}
\label{appendix_A}

Results from the Monte Carlo study described in section \ref{MC} were cross-checked on toy Monte Carlo data sets. Samples of $\sim600$ signal and background particles were generated according to $\eta$ and $p_T$ distributions similar to those obtained using Pythia. Particle $\eta$ and $p_T$ were generated independently: Gaussian PDFs centered at zero with standard deviations comparable to those observed in Monte Carlo were used for $\eta$, and $p_T$ values were generated based on polynomial PDFs in the range $2~\mbox{GeV/c}<p_T<5~\mbox{GeV/c}$ parametrizing the corresponding Monte Carlo distributions.

Additional cross-checks were performed by varying the toy Monte Carlo generation parameters by $\pm 10\%$ with respect to the nominal values, in order to make sure that the 
results did not depend on the specific parameter choice. The algorithm was also run on different numbers of particles in the input data set, with no appreciable changes to the results.

\newpage


\end{multicols}
\end{document}